\documentclass[aip,cha,reprint]{revtex4-1}

\usepackage[usenames,dvipsnames]{xcolor}
\usepackage{amsmath}
\usepackage{amssymb,mathrsfs}
\usepackage{graphicx}
\usepackage{mathtools}

\newcommand{\prt}{\partial}
\newcommand{\la}{\lambda}

\newcommand{\sn}{\mathrm{sn}}
\newcommand{\cn}{\mathrm{cn}}

\usepackage{hyperref}
\allowdisplaybreaks

\begin{document}

\title{Evolution of initial discontinuities in the Riemann problem for the
Kaup-Boussinesq equation with positive dispersion}

\author{T. Congy}\affiliation{LPTMS, CNRS, Univ. Paris-Sud,
Universit\'e
Paris-Saclay, 91405 Orsay, France}

\author{S. K. Ivanov}\affiliation{Institute of Spectroscopy, Russian
  Academy of Sciences, Troitsk, Moscow, 108840, Russia}
\affiliation{Moscow Institute of Physics and Technology, Institutsky
  lane 9, Dolgoprudny, Moscow region, 141700, Russia}

\author{A. M. Kamchatnov}\affiliation{Institute of Spectroscopy,
  Russian Academy of Sciences, Troitsk, Moscow, 108840, Russia}

\author{N. Pavloff}\affiliation{LPTMS, CNRS, Univ. Paris-Sud, Universit\'e
Paris-Saclay, 91405 Orsay, France}

\begin{abstract}
  We consider the space-time evolution of initial discontinuities of
  depth and flow velocity for an integrable version of the shallow
  water Boussinesq system introduced by Kaup. We focus on a specific
  version of this ``Kaup-Boussinesq model'' for which a flat water
  surface is modulationally stable, we speak below of ``positive
  dispersion'' model. This model also appears as an approximation to
  the equations governing the dynamics of polarisation waves in
  two-component Bose-Einstein condensates. We describe its periodic
  solutions and the corresponding Whitham modulation equations. The
  self-similar, one-phase wave structures
  are composed of different building blocks which are studied in
  detail. This makes it possible to establish a classification of all
  the possible wave configurations evolving from initial discontinuities.
  The analytic results are confirmed by numerical simulations.
\end{abstract}



\maketitle

{\bf The Kaup-Boussinesq water wave equation is integrable, but its
  generic form suffers from a dynamical instability. We study here the
  Riemann problem for a version of this equation which does not suffer
  from the same deficiency. This equation appears as an approximation
  of the nonlinear polarization dynamics of a two-component
  Bose-Einstein condensate, and in this context it is important to
  characterize the time evolution of simple, but experimentally
  relevant initial profiles.}

\section{Introduction}

In many physical wave systems, the initial value problem treated in
the long-wavelength (hydrodynamic) approximation leads to wave
breaking after a finite time. As a result, the formal solution becomes
multivalued, i.e. it looses its physical meaning. At the wave breaking
point, the first spatial derivative of the physical variables diverges
and the hydrodynamic approximation fails. This suggests that this
nonphysical behavior can be remedied by accounting for physical
effects described by terms with higher-order derivatives in the
corresponding evolution equations. For example, within the
Navier-Stokes description of the dynamics of a compressible gas, the
effects of viscosity are described by second-order derivative terms,
and this leads, instead of wave breaking, to the formation of viscous
shocks, which can often be formally described by surfaces of
discontinuities  in the physical variables. Formulated
in this way, the theory of ``shock waves'' has found a number of
important applications \cite{Cou56,Zel66}.

At the same time, in many physical systems the dissipative effects may
be relatively weaker than the dispersive ones, and in such cases,
so-called ``dispersive shock waves'' (DSWs) are formed instead of
viscous shocks. DSWs can be represented as modulated nonlinear
oscillations whose envelope varies over characteristic distances much
greater than their wavelength. In recent years such systems have
attracted much attention in fluid dynamics, nonlinear optics, physics
of Bose-Einstein condensates and other areas of physics (see, e.g.,
Refs.  \onlinecite{Kam00,eh-16}). This type of problem was studied for
the first time in the context of the physics of shallow water waves
whose evolution is described by the celebrated Korteweg-de Vries (KdV)
equation \cite{kdv-1895,bl-54}.  The equations governing the slow
evolution of the envelope of the nonlinear oscillations had been
derived by Whitham \cite{whitham-65} and later they were applied to the
description of the DSW structure by Gurevich and Pitaevskii
\cite{gp-73}. Because of the universality of the KdV equation, this
approach can naturally be applied to many other physical situations.
When the condition of unidirectional propagation is relaxed, shallow
water waves are described by various forms of Boussinesq equation
\cite{boussinesq-1877}. The most convenient form for our purpose has
been derived by Kaup \cite{kaup-75}; this is the so-called
Kaup-Boussinesq (KB) equation. The KB equation is completely
integrable and the well-developed methods of inverse scattering
transform and finite-gap integration can be used for explicitly
deriving its multi-soliton and (quasi-)periodic solutions
\cite{my-79}.

In the applications of this theory to concrete physical problems, only
the KB equations with negative dispersion has been considered so far.
In this case, linear perturbations can be sought under the form of
plane waves with angular frequency $\omega$ and wavelength $k$. In
non-dimensional notations, the corresponding dispersion relation reads
($h_0$ is a constant depth)
\begin{equation}\label{eq1}
\omega^2=h_0k^2-\frac14k^4.
\end{equation}
The Whitham modulation equations were derived for this case in
Ref.~\onlinecite{egp-01}, and a complete classification of all the
possible wave structures resulting from an initial discontinuous
profile were obtained. Besides that, the analytic solution for a
generic wave breaking regime was found in Ref.~\onlinecite{egk-05a} --
with the use of a generalized hodograph transform -- and stationary
undular bore structures whose form was stabilized by weak viscous
effects were studied in Ref.~\onlinecite{egk-05b}. However, the
dispersion relation \eqref{eq1} corresponds to a dynamical instability
of small wavelength perturbations over a fluid of constant depth
$h_0$. There exists another form of the KB system, with positive
dispersion, for which the dispersion relation of linear waves reads
\begin{equation}\label{eq2}
\omega^2=h_0k^2+\frac14k^4.
\end{equation}
The corresponding KB system can be written under the following
non-dimensional form,
\begin{equation}\label{eq3}
 \begin{array}{l}
  h_t+(hu)_x-\frac14u_{xxx}=0,\\
  u_t+uu_x+h_x=0.
 \end{array}
\end{equation}
In the context of shallow water wave physics, $h$  is the local
height of the water layer and $u$ is a local mean flow
velocity\cite{rem-KB}.  Eq.~(\ref{eq2}) represents the dispersion
relation of linear waves propagating along a uniform background
characterized by the physical variables $h_0=\mathrm{const}$ and
$u_0=0$. It does not suffer from the instability of Eq. \eqref{eq1}.
The positive dispersion KB system \eqref{eq3} may be obtained in the
case of capillary waves propagating on top of a thin fluid layer (see,
e.g., Ref. \onlinecite{Kam00}). Besides this physical realization, the
system (\ref{eq3}) appears as an approximation to the Landau-Lifshitz
equation for propagation of magnetization waves in easy-plane magnets
and to the Gross-Pitaevskii equations for propagation of polarization
waves in two-component Bose-Einstein condensates \cite{Iva17}.

Motivated by these applications of the KB system (\ref{eq3}), we
consider  in the present paper the so-called Riemann problem. This
corresponds to the study of the time evolution of initial
discontinuous profiles of the form
\begin{equation}\label{eq4}
 \begin{array}{l}
  h(x,t=0)=h_L,\;\;\text{and}\;\; u(x,t=0)=u_L\;\;\text{for}\;\; x<0,\\
  h(x,t=0)=h_R,\;\;\text{and}\;\;u(x,t=0)=u_R\;\;\text{for}\;\; x>0,
 \end{array}
\end{equation}
As we shall see, the resulting wave structures differ considerably
from those found in Ref.~\onlinecite{egp-01} for the negative dispersion
case. In the case of Eqs.~(\ref{eq3}) studied in the present work,
the classification of the
possible wave structures follows closely the scheme found for the
nonlinear Schr\"{o}dinger equation in Refs.~\onlinecite{gk-87,eggk-95}. We
shall obtain simple analytic formulae for the main parameters of the
wave structures and confirm their accuracy by comparison with
numerical solutions of the KB system \eqref{eq3}.

\section{Periodic waves and Whitham modulation
  equations}\label{sec.whitham}

In this section we derive the periodic wave solutions (the so called
cnoidal waves) of the system \eqref{eq3} and the Whitham equations
governing the modulational dynamics of a cnoidal wave. This is
achieved by using the methods described, e.g., in Ref. \onlinecite{Kam00} (see
also \onlinecite{egk-05a}). These techniques are based on the possibility to
represent the system (\ref{eq3}) as a compatibility condition for the
linear system \cite{kaup-75}
\begin{equation}\label{eq5}
\psi _{xx}=\mathcal{A}\, \psi ,
\qquad
\psi _{t}=-\frac12\mathcal{B}_x\psi+\mathcal{B}\,\psi_x
\end{equation}
with
\begin{equation}\label{eq6}
  \mathcal{A}=h-\left(\la-\frac12 u\right)^2,\quad\mbox{and}\quad
  \mathcal{B}=-\left(\la+\frac12 u\right),
\end{equation}
where $\la$ is a free spectral parameter. Demanding that
$(\psi_{xx})_t=(\psi_t)_{xx}$ for any $\la$, we reproduce the KB
system (\ref{eq3}).

The second order spatial linear differential equation in (\ref{eq5})
has two independent solutions $\psi _{+}(x,t)$ and $\psi
_{-}(x,t)$. Their product $g=\psi _{+}\psi _{-} $ satisfies the
following third order equation
\begin{equation}\label{eq7}
g_{xxx}-2\mathcal{A}_xg-4\mathcal{A}\, g_x=0.
\end{equation}
Upon multiplication by $g$, this equation can be integrated once to
give
\begin{equation}\label{eq8}
\frac12 gg_{xx}-\frac14 g_x^2-\mathcal{A}g^2=P(\la),
\end{equation}
where the integration constant has been written as $P(\lambda )$ since
it can only depend on $\lambda$.  The time dependence of $g(x,t)$ is
determined by the equation
\begin{equation}\label{eq9}
g_{t}=\mathcal{B}\,g_x-\mathcal{B}_xg.
\end{equation}
We are interested in the one-phase periodic solution of the system
(\ref{eq3}). They are distinguished by the condition that $P(\lambda
)$ in (\ref{eq8}) be a fourth degree polynomial of the
form\cite{defs}
\begin{equation}\label{eq10}
P(\lambda )=\prod_{i=1}^{4}(\lambda -\lambda _{i})=\lambda^{4}-
s_{1}\lambda^{3}+s_{2}\lambda ^{2}-s_{3}\lambda +s_{4}.
\end{equation}
In expression \eqref{eq10}, we chose for definiteness to order
the zeroes $\la_i$ according to
\begin{equation}\label{eq10a}
 \la_1\leq\la_2\leq\la_3\leq\la_4.
\end{equation}
Then we find from Eq.~(\ref{eq8}) that $g(x,t)$ is a first-degree
polynomial in $\lambda$, of the form
\begin{equation}\label{eq11}
g(x,t)=\lambda -\mu (x,t),
\end{equation}
where $\mu (x,t)$ is connected with $u(x,t)$ and $h(x,t)$ by the
relations
\begin{equation}\label{eq12}
\begin{split}
& u(x,t)=s_{1}-2\, \mu (x,t),\\
& h(x,t)=\tfrac{1}{4}s_{1}^{2}-s_{2}-2\mu^{2}(x,t)
+s_{1}\mu(x,t) ,\end{split}
\end{equation}
which follow from a comparison of the coefficients of the different
powers of $\lambda$ on both sides of Eq.~(\ref{eq8}). The spectral
parameter $\lambda $ is arbitrary and on substitution of $\lambda =\mu
$ into Eq.~(\ref{eq8}) we obtain an equation for $\mu $,
\[
\mu _{x}=2\sqrt{- P(\mu )},
\]
while a similar substitution into Eq.~(\ref{eq9}) gives
\[
\mu _{t}=-(\mu +\tfrac{1}{2}u)\mu _{x}=-\tfrac{1}{2}s_{1}\mu _{x}.
\]
Hence, $\mu (x,t)$ as well as $u(x,t)$ and $h(x,t)$ depend only on the
phase
\begin{equation}\label{eq13}
\theta =x-\tfrac{1}{2}s_{1}t,
\end{equation}
so that
\begin{equation}\label{eq14}
  V=\frac12 s_1=\frac12\sum_{i=1}^4\la_i
\end{equation}
is the phase velocity of the nonlinear wave,
and $\mu (\theta )$ is determined by the equation
\begin{equation}\label{eq15}
\mu _{\theta }=2\sqrt{ -P(\mu )}.
\end{equation}

It follows from Eq.~(\ref{eq12}) that the variable $\mu$ must be real.
For the fourth degree polynomial (\ref{eq10}) the real solution of
\eqref{eq15} corresponds to oscillations of $\mu$ in one of two
possible intervals,
\begin{equation}\label{eq16}
 \la_1\leq\mu\leq\la_2\qquad \text{or}\qquad \la_3\leq\mu\leq\la_4,
\end{equation}
within which $P(\mu)$ assumes negative values. It is well known that
the solution of Eq. \eqref{eq15} with boundaries (\ref{eq16}) can be
expressed in terms of elliptic functions (see, e.g,, Refs.
\onlinecite{gardner-2012,ckp-16}). Without going into details, we shall
list here the results which are the most relevant to our study.

$\bullet$ For the case
\begin{equation}\label{eq17}
    \la_1\leq\mu\leq\la_2
\end{equation}
the cnoidal wave solution of Eq.~(\ref{eq16}) with the initial condition
$\mu(0)=\la_1$ is given by
\begin{equation}\label{eq18}
\mu(\theta)=\la_2-
\frac{(\la_2-\la_1)
\cn^2\left(W,m\right)}
{1+\frac{\la_2-\la_1}{\la_4-\la_2}
\sn^2\left(W, m \right)},
\end{equation}
where $W=\sqrt{(\la_3-\la_1)(\la_4-\la_2)}\,\theta$ and
\begin{equation}\label{eq19}
    m=\frac{(\la_2-\la_1)(\la_4-\la_3)}{(\la_3-\la_1)(\la_4-\la_2)}
\end{equation}
is the modulus of the Jacobi elliptic functions $\sn$ and $\cn$.
Substitution of (\ref{eq18}) into (\ref{eq12}) gives the corresponding
expressions for $u(\theta)$ and $h(\theta)$ for a one-phase periodic
nonlinear wave. Its wavelength is given by
\begin{equation}\label{eq20}
    L=\int_{\la_1}^{\la_2}\frac{d\mu}{\sqrt{-P(\mu)}}=
    \frac{2K(m)}{\sqrt{(\la_3-\la_1)(\la_4-\la_2)}},
\end{equation}
$K(m)$ being the complete elliptic integral of the first kind.  The
soliton solution corresponds to the limit $\la_3\to\la_2$
$(m\to1)$. We obtain
\begin{equation}\label{eq21}
    \mu(\theta)=\la_2-\frac{\la_2-\la_1}
{\cosh^2 W+
\frac{\la_2-\la_1}{\la_4-\la_2}
\sinh^2 W}.
\end{equation}
This is a dark soliton solution for the variable $\mu$. In the limit
$\la_2\to\la_1$ we get a small-amplitude harmonic wave
\begin{equation}\label{eq22}
\begin{split}
&  \mu=\la_2-\frac12(\la_2-\la_1)\cos[k(x-Vt)],\\
\qquad\mbox{where}\quad &
  k=2\sqrt{(\la_3-\la_1)(\la_4-\la_1)}.\end{split}
\end{equation}
If $\la_4=\la_3$ but $\la_1\neq\la_2$, then we have again $m=0$ and
(\ref{eq21}) reduces to a nonlinear trigonometric wave, but we shall
not present its explicit form here (cf., e.g., Refs.
\onlinecite{gardner-2012,ckp-16}).

$\bullet$ In a similar way, for the case
\begin{equation}\label{eq22a}
    \la_3\leq\mu\leq\la_4,
\end{equation}
the cnoidal wave solutions are of the form ($\mu(0)=\la_4$)
\begin{equation}\label{eq23}
\mu(\theta)=\la_3+\frac{(\la_4-\la_3)
\cn^2\left(W,m\right)}
{1+\frac{\la_4-\la_3}{\la_3-\la_1}
\sn^2\left(W, m \right)}.
\end{equation}
In the soliton limit $\la_3\to\la_2$ $(m\to1)$ we obtain
\begin{equation}\label{eq24}
    \mu(\theta)=\la_2+\frac{\la_4-\la_2}
{\cosh^2W+
\frac{\la_4-\la_2}{\la_2-\la_1}
\sinh^2W}.
\end{equation}
This is a bright (for $\mu$-variable) soliton over a constant background.
In the limit $\la_4\to\la_3$ we get a small-amplitude harmonic wave
\begin{equation}\label{eq25}
\begin{split}
& \mu=\la_3+\frac12(\la_4-\la_3)\cos[k(x-Vt)],\\
\qquad\mbox{where}\quad  &
k=2\sqrt{(\la_3-\la_1)(\la_3-\la_1)}\; .
\end{split}
\end{equation}
As discussed above, nonlinear trigonometric waves also exit, here in
the case where $\la_1=\la_2$ but $\la_3\neq\la_4$. If furthermore
$\la_3\to\la_1$ one reaches the limit of an algebraic
soliton\cite{gardner-2012,ckp-16}:
\begin{equation}\label{alg-sol}
\mu(\theta)=\la_1+\frac{\la_4-\la_1}{1+(\la_4-\la_1)^2\theta^2}\; .
\end{equation}

$\bullet$ We now consider slowly modulated cnoidal waves. In this case the
parameters $\la_i$ ($i=1,2,3,4$) become slowly varying functions of
$x$ and $t$ changing weakly over a wavelength $L$. Their evolution is
governed by the Whitham modulation equations \cite{Kam00,eh-16}
\begin{equation}\label{eq26}
    \frac{\prt\la_i}{\prt t}+v_i\, \frac{\prt\la_i}{\prt x}=0,
    \quad i=1,2,3,4.
\end{equation}
The  Whitham velocities $v_i$ appearing in Eqs. \eqref{eq26}
can be computed {\it via} the formulae (see, e.g., \cite{Kam00,eh-16})
\begin{equation}\label{eq27}
    v_i(\la_1,\la_2,\la_3,\la_4)
=\left(1-\frac{L}{\partial_{\lambda_i}L}\prt_{\lambda_i}\right)V,
\quad i=1,2,3,4,
\end{equation}
where the phase velocity $V$ and the wavelength $L$ are given
by Eqs.~(\ref{eq14}) and (\ref{eq20}). A simple
calculation yields the explicit expressions
\begin{equation}\label{eq28}
    \begin{array}{l}
\displaystyle{
    v_1=\frac12\sum_{i=1}^4\la_i-\frac{(\la_4-\la_1)(\la_2-\la_1)K(m)}
    {(\la_4-\la_1)K(m)-(\la_4-\la_2)E(m)},}\\[5mm]
\displaystyle{
    v_2=\frac12\sum_{i=1}^4\la_i+\frac{(\la_3-\la_2)(\la_2-\la_1)K(m)}
    {(\la_3-\la_2)K(m)-(\la_3-\la_1)E(m)},}\\[5mm]
\displaystyle{
    v_3=\frac12\sum_{i=1}^4\la_i-\frac{(\la_4-\la_3)(\la_3-\la_2)K(m)}
    {(\la_3-\la_2)K(m)-(\la_4-\la_2)E(m)},}\\[5mm]
\displaystyle{
    v_4=\frac12\sum_{i=1}^4\la_i+\frac{(\la_4-\la_2)(\la_4-\la_1)K(m)}
    {(\la_4-\la_1)K(m)-(\la_3-\la_1)E(m)},}
    \end{array}
\end{equation}
where $m$ is given by \eqref{eq19} and
$K(m)$ and $E(m)$ are complete elliptic integrals of
the first and second kind, respectively.

In the soliton limit $m \to 1$ (i.e., $\la_3 \to \la_2$) the Whitham
velocities reduce to
\begin{equation}\label{eq29}
   \begin{split}
&    v_1=\frac12(3\la_1+\la_4),\quad v_2=v_3=\frac12(\la_1+2\la_2+\la_4),\\
& v_4=\frac12(\la_1+3\la_4).
    \end{split}
\end{equation}
In a similar way, in the small amplitude limit $m\to0$ (i.e.,
$\la_2\to\la_1$) we obtain
\begin{equation}\label{eq30}
\begin{split}
& v_1=v_2=2\la_1+\frac{(\la_4-\la_3)^2}{2(\la_3+\la_4-2\la_1)},\\
& v_3=\frac12(3\la_3+\la_4),\quad v_4=\frac12(\la_3+3\la_4),
\end{split}
\end{equation}
and in another small amplitude limit ($m\to0$ when $\la_3\to\la_4$) we have
\begin{equation}\label{eq31}
\begin{split}
&    v_1=\frac12(3\la_1+\la_2),\quad v_2=\frac12(\la_1+3\la_2),\\
&    v_3=v_4=2\la_4+\frac{(\la_2-\la_1)^2}{2(\la_1+\la_2-2\la_4)}.
   \end{split}
\end{equation}

\section{Key elements of self-similar wave structures}

The initial profiles (\ref{eq4}) being infinitely sharp, do not
involve any characteristic length. However the dispersion relation
(\ref{eq2}) is characterized by the value of the shallow water wave
velocity: $c_s=\left.\omega/k\right|_{k\to0}=\sqrt{h_0}$. Therefore
the large scale features of the solution of this problem (with
characteristic length scale much greater than the wavelength) can
only depend on the self-similar variable $\xi=x/t$, which can be made
non-dimensional with the help of the velocity $c_s$. This means that
the large scale features of
the wave pattern must be self-similar and should
be composed of (possibly several) regions
where $h$ and $v$ either smoothly depend on $\xi$, or consist of
modulated periodic waves whose envelopes (and wavelength $L$)
depend slowly on $\xi$.

In the framework of the hydrodynamic approximation these regions are
separated by weak discontinuities where the physical variables have
cusps. If the hydrodynamic approximation leads to non-monotonous dependence of
velocities on the wave amplitude,
then the wave structure can be more complicated; an example of such a
situation was considered, e.g., in Ref.~\onlinecite{gardner-2012}.
At first we shall consider smooth solutions of the KB
system (\ref{eq3}).

\subsection{Dispersionless limit}

For smooth enough wave patterns we can neglect the last dispersion term in
the first equation of the system (\ref{eq3}) and arrive at the so-called
dispersionless equations
\begin{equation}\label{eq29a}
h_t+(hu)_x=0,\qquad u_t+uu_x+h_x=0,
\end{equation}
which coincide with the well-known shallow water equations. Introducing
the Riemann invariants
\begin{equation}\label{eq31a}
\la_{\pm}=\frac{u}2\pm\sqrt{h}\;,
\end{equation}
the system \eqref{eq29a} can be
written in the following diagonal form
\begin{equation}\label{eq30a}
\begin{split}
& \frac{\prt\la_{\pm}}{\prt t}+v_\pm(\la_-,\la_+)\frac{\prt\la_\pm}{\prt x}=0,\\
& \quad\mbox{where}\quad
v_\pm(\la_-,\la_+)=\frac12(3\la_\pm+\la_\mp)\; .
\end{split}
\end{equation}
The physical variables are expressed in terms of $\la_{\pm}$ as
\begin{equation}\label{eq31b}
u=\la_++\la_-,\qquad h=(\la_+-\la_-)^2/4.
\end{equation}
For the self-similar solutions one has $\la_{\pm}=\la_{\pm}(\xi)$ and
the system (\ref{eq30a}) reduces to
\begin{equation}\label{eq32}
\frac{d\la_+}{d\xi}\cdot\left(v_+-\xi\right)=0,\qquad
\frac{d\la_-}{d\xi}\cdot\left(v_--\xi\right)=0.
\end{equation}
This system admits a trivial solution for which $\la_+=\mathrm{const}$ and
$\la_-=\mathrm{const}$. It describes a uniform flow
with constant $h$ and $u$. We shall call
such a solution a ``plateau''.

\begin{figure}
\begin{center}
\includegraphics[width=0.99\linewidth]{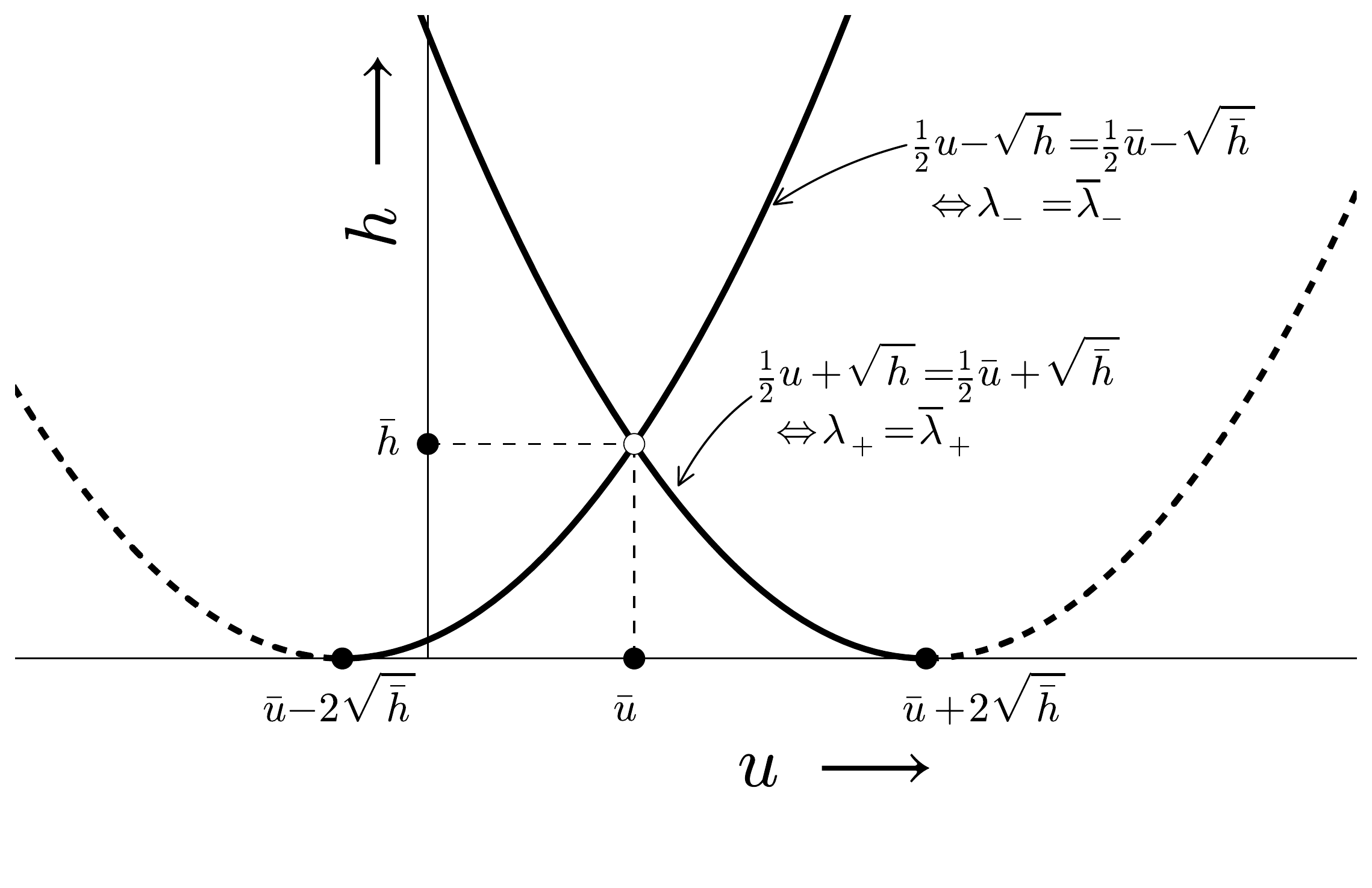}
\end{center}
\caption{Relation between $u$ and $h$ for simple wave solutions in the
  dispersionless regime. The solid lines correspond to the curve
  $\lambda_+=C^{\rm st}=\overline{\lambda}_+$ [portion of parabola
  ending at point with coordinates $(\bar{u}+2\sqrt{\bar{h}},0)$] and
  to the curve $\lambda_-=C^{\rm st}=\overline{\lambda}_-$ [portion of
  parabola ending at point $(\bar{u}-2\sqrt{\bar{h}},0)$].  They are
  continued by the dashed curves along which
  $\lambda_-=\overline{\lambda}_+$ (right dashed curve) and
  $\lambda_+=\overline{\lambda}_-$ (left dashed curve)
  respectively. These dashed curves are of no significance for the
  present discussion of simple waves, but will become important in
  Sec. \ref{sec.classif}.}
\label{fig1}
\end{figure}
Other solutions of \eqref{eq32} are called simple waves.  For
such flows, one of the Riemann invariants is constant (say,
$\lambda_-$) whereas the other one changes in such a way that the term
between parenthesis in its equation is zero ($v_+=\xi$ in the example
considered). One has thus two possible types of self-similar simple waves:
\begin{equation}\label{eq32a}
\begin{cases}
\lambda_-=C^{\rm st}\equiv\overline{\lambda}_-\; ,
\quad \mbox{with}\\
v_+(\overline{\la}_-,\la_+)=
\frac12(3\la_++\overline{\la}_-)=\xi=x/t\; ,
\end{cases}
\end{equation}
or
\begin{equation}\label{eq32b}
\begin{cases}
 \lambda_+=C^{\rm st}\equiv\overline{\lambda}_+\; ,
\quad \mbox{with}\\
v_-(\la_-,\overline{\la}_+)=
\frac12(3\la_-+\overline{\la}_+)=\xi=x/t\; .
\end{cases}
\end{equation}
The constancy of one of the Riemann invariants means that $h$ and $u$
are related by a simple formula: either $\la_-=
u/2-{h}^{1/2}=\mathrm{const}=\overline{\la}_-= \bar{u}/2-\bar{h}^{1/2} $, or
$\la_+= u/2+{h}^{1/2}=\mathrm{const}=\overline{\la}_+=
\bar{u}/2+\bar{h}^{1/2} $, where $\bar{u}$ and $\bar{h}$ are some
values that fix the value of the constant Riemann invariant and can be
chosen at convenience for solving a specific problem.  Thus, for given
values of $\bar{u}$ and $\bar{h}$, a simple wave corresponds to a
configuration where the variables $u$ and $h$ are connected by the
relations corresponding to one of the two parabolas drawn in the plane
$(u,h)$ in Fig.~\ref{fig1}.  The parabolae cross at the point
$P=(\bar{u},\bar{h})$ which represents a uniform flow with constant
values of $u=\bar{u}$ and $h=\bar{h}$ which is a trivial ``plateau
solution'' of Eqs.~(\ref{eq29a}).

\begin{figure}
\begin{center}
\includegraphics[width=0.99\linewidth]{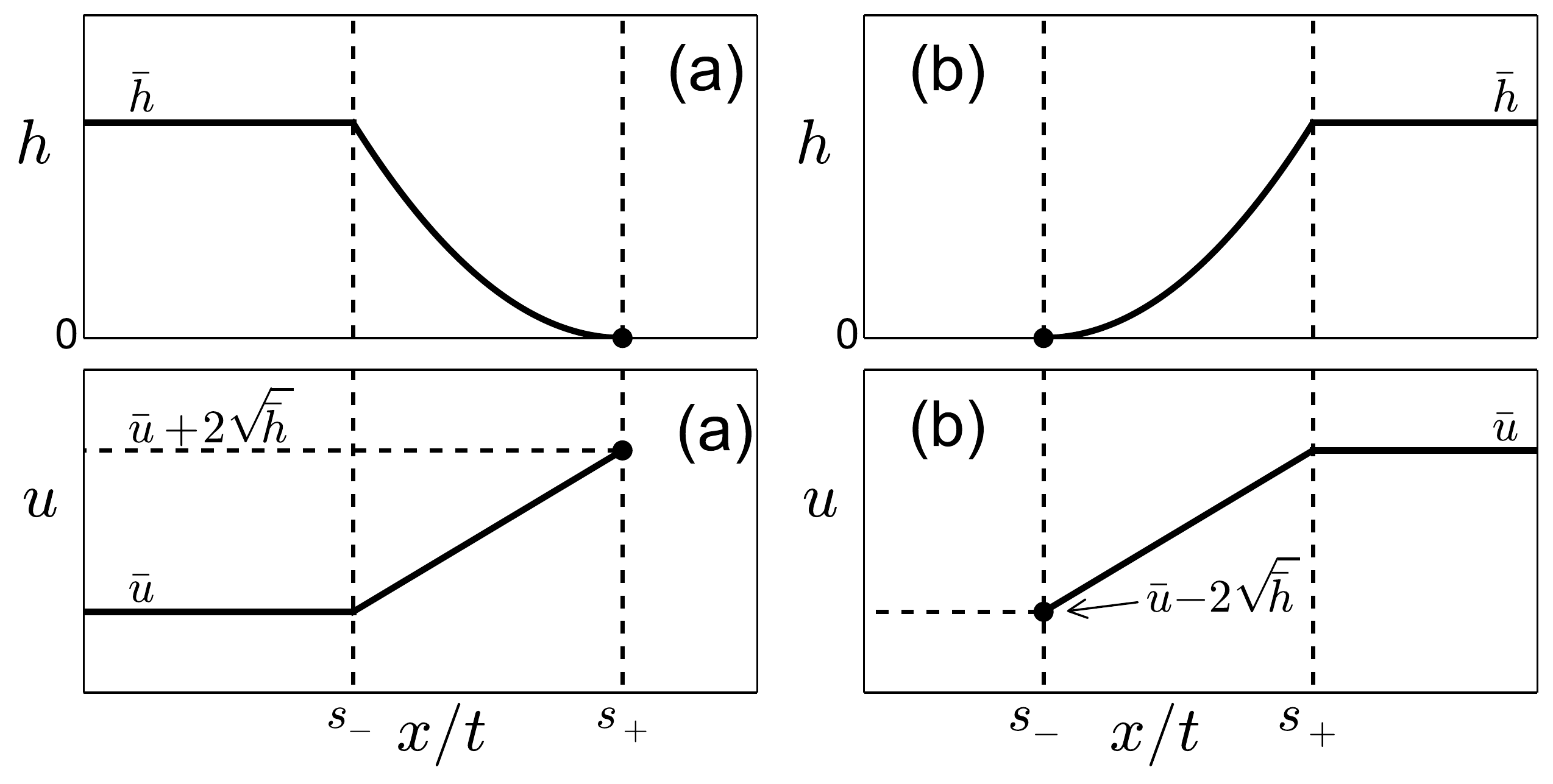}
\end{center}
\caption{Height and velocity profiles for self-similar rarefaction
  wave solutions of the dispersionless equations \eqref{eq29a}
  expanding into empty space (the so called ``dam
  problem''). Figs. (a) corresponds to a flow expanding in the
  positive $x$ direction and Figs. (b) to a flow expanding in the
  negative $x$ direction. The values of the edge velocities $s_\pm$
  are given in the text.}
\label{fig2}
\end{figure}

The dispersionless system (\ref{eq29a}) requires continuity of the
functions $u(x,t)$ and $h(x,t)$, but, as usual in hydrodynamics,
admits jumps of their space derivatives, i.e., ``weak
discontinuities''. Therefore a plateau solution $(\bar{u},\bar{h})$
can be attached at one of its boundaries to a simple wave. We have
here two possibilities. If the uniform flow corresponding to
$P=(\bar{u},\bar{h})$ matches with the simple wave along which
$\la_+=\overline{\la}_+=\mathrm{const}$ (one of the
solid parabolas in
Fig.~\ref{fig1}), then from Eq. \eqref{eq32b}
one gets for this flow
\begin{equation}\label{eq33}
\begin{cases}
u(x,t)=\frac23\left(\frac{x}t+\overline{\la}_+\right)
=\frac23\left(\frac{x}t+\frac{\bar{u}}2
+\sqrt{\bar{h}}\,\right),\\
h(x,t)=\frac19\left(\frac{x}t-2\overline{\la}_+\right)
=\frac19\left(\frac{x}t-\bar{u}-2\sqrt{\bar{h}}\,\right)^2.
\end{cases}
\end{equation}
This wave configuration represents a rarefaction wave (RW) propagating to
the right. If it propagates into `vacuum', then Eq.~(\ref{eq33}) gives
the full solution of the problem (\ref{eq4}) with left boundaries
$h_L=\bar{h}$ and $u_L=\bar{u}$ whereas at the right boundary
$h_R=0$ (the value of $u_R$ is
irrelevant in the space without fluid). This situation is depicted in
Figs.~\ref{fig2}(a).  The left edge of this rarefaction wave propagates
to the left at velocity $s_-={u_L}-\sqrt{{h}_L}$ and the right
edge propagates to the right into the empty space with velocity
$s_+=u_L+2\sqrt{{h}_L}$.

In a similar way, if $P=(\bar{u},\bar{h})$ matches with the simple wave along
which $\la_-=\overline{\la}_-=\mathrm{const}$, then \eqref{eq32a} yields
\begin{equation}\label{eq34}
\begin{cases}
u(x,t)=\frac23\left(\frac{x}t+\overline{\la}_-\right)
=\frac23\left(\frac{x}t+\frac{\bar{u}}2+\sqrt{\bar{h}}\,\right),
\\
h(x,t)=\frac19\left(\frac{x}t-2\overline{\la}_-\right)
=\frac19\left(\frac{x}t-\bar{u}+2\sqrt{\bar{h}}\,\right)^2.
\end{cases}
\end{equation}
This represents a rarefaction wave propagating to the left. Again, it
corresponds---in the hydrodynamic approximation---to the solution of
the problem (\ref{eq4}) with $h_L=0$ whereas at the right boundary
$h_R=\bar{h}$ and $u_R=\bar{u}$, see Figs.~\ref{fig2}(b). The edge
velocities are equal to $s_-={u_R}-2\sqrt{h_R}$ and
$s_+={u_R}+\sqrt{h_R}$.

\begin{figure}
\begin{center}
\includegraphics[width=0.99\linewidth]{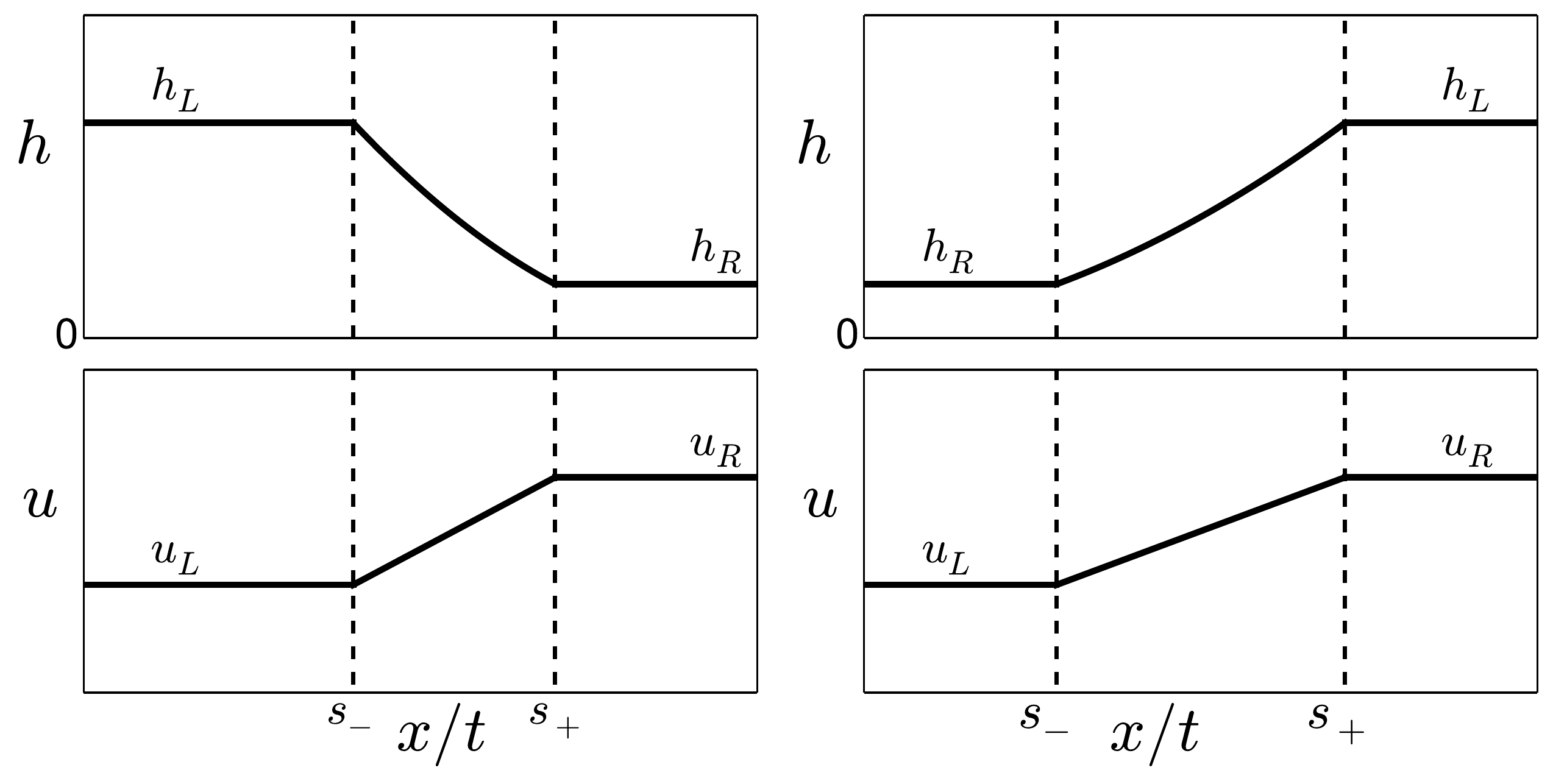}
\end{center}
\caption{Self-similar solutions of the dispersionless
  equations \eqref{eq29a} composed by a rarefaction wave
connecting two uniform flows. The two left
  plots correspond to the situation where $\lambda_+=$ const for the
  whole flow, the two right ones to $\lambda_-=$ const.}
\label{fig3}
\end{figure}

It is clear that we can generalize these solutions to the cases where
both sides of the rarefaction wave connect uniform flows with equal
values of the corresponding Riemann invariants ${u_L}/2+\sqrt{{h}_L}=
{u}_R/2+\sqrt{{h}_R}$ or ${u_L}/2-\sqrt{{h}_L}=
{u}_R/2-\sqrt{{h}_R}$. In these cases the rarefaction wave connects
two uniform flows and the corresponding distributions of $h$ and $u$
are shown in Fig.~\ref{fig3}. The velocities of the edges of the
rarefaction waves are given in both cases by the formulae
$s_-=u_L-\sqrt{h_L}$, $s_+=u_R+\sqrt{h_R}$. These values have simple
physical interpretation: they are the sums of the local flow
velocities ($u_L$ or $u_R$) and of the propagation velocities of small
amplitude disturbances directed to the left for the left edge
($-\sqrt{h_L}$) and to the right for the right edge ($+\sqrt{h_R}$).

\begin{figure}
\begin{center}
\includegraphics[width=0.99\linewidth]{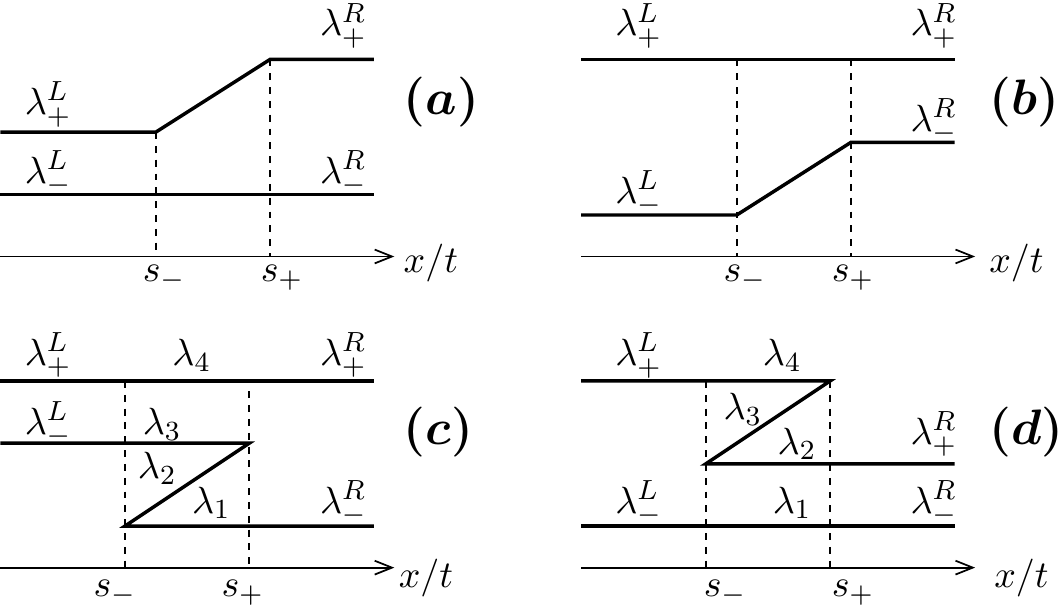}
\end{center}
\caption{Diagrams representing the evolution of the Riemann invariants
  as a function of $x/t$. The plots (a) and (b) correspond to the
  configuration where a dispersionless rarefaction wave connects two
  uniform flows. Plots (c) and (b) considered within the
  dispersionless approximation correspond to a formal multi-valued
  solution. In this case, the dispersionless approximation breaks down
  and one observes a dispersive shock wave, accurately described by 4
  Riemann invariants within the Whitham modulational approach,
  cf. Sec. \ref{sec.DSW}.}
\label{fig4}
\end{figure}

It is important to note that the system (\ref{eq32}) only admits
solutions of the type (\ref{eq32a}) and (\ref{eq32b}) for which the
non-constant Riemann invariant increases with $\xi=x/t$. The above
wave structures correspond to the conditions (a) $\la_+^L<\la_+^R$,
$\la_-^L=\la_-^R$ or (b) $\la_+^L = \la_+^R$, $\la_-^L<\la_-^R$, as
illustrated in Figs.~\ref{fig4}(a,b). The other two situations
represented in Figs.~\ref{fig4}(c,d), that is (c) $\la_+^L = \la_+^R$,
$\la_-^L > \la_-^R$ and (d) $\la_+^L > \la_+^R$, $\la_-^L = \la_-^R$,
result in multi-valued solutions and are therefore nonphysical: the
dispersionless approximation is not applicable in these cases and we
have to turn to another type of key elements for describing such
structures.

\subsection{Dispersive shock waves}\label{sec.DSW}

Since the pioneering work of Gurevich and Pitaevskii \cite{gp-73}, it
is known that wave breaking---such as depicted in
Figs.~\ref{fig4}(c,d) for the dispersionless Riemann invariants---is
regularized by the replacement of the nonphysical multi-valued
dispersionless solution by a dispersive shock wave. This wave pattern
can be represented approximately as a modulated nonlinear periodic
wave whose parameters $\la_i$ ($i=1,2,3,4$,
cf. Sec.~\ref{sec.whitham}) change slowly along the wave structure. In
this case, the two dispersionless Riemann invariants $\la_{\pm}$ are
replaced in the DSW region by four Riemann invariants $\la_i$,
cf. Fig.~\ref{fig4}(c,d).  In this region, the evolution of the DSW is
determined by the Whitham equations (\ref{eq26}).  In our case, when we
consider of self-similar solution, all Riemann invariants depend only
on $\xi=x/t$, and the Whitham equations reduce to
\begin{equation}\label{eq35}
 \frac{d\la_i}{d\xi}\cdot
\left[v_i(\la_1,\la_2,\la_3,\la_4)-\xi\right]=0,\quad i=1,2,3,4.
\end{equation}
One of the two factors in this equation must vanish, that is, for each
$i$, either the Riemann invariant $\la_i$ is constant, or
$v_i=\xi$. For the Whitham velocities (\ref{eq28}) there exist
solutions for which, in the DSW region, only one of the Riemann
invariants vary, whereas the three others remain constant.  These
solutions correspond qualitatively to the same patterns as the ones
depicted in Figs.~\ref{fig4}(c,d). However, naturally, the dependence
of the Riemann invariants on $\xi$ resulting from \eqref{eq35} differs
from the one obtained from \eqref{eq32} (different equations and
different variables). As a result, the velocities of the edges of the
DSW do not coincide with the velocities of the nonphysical solutions
of the dispersionless equations. It nonetheless remains true that
Figs.~\ref{fig4}(c,d) schematically represent the structure of the
Riemann invariants for the solutions of the Whitham equations
(\ref{eq35}): in Fig.~\ref{fig4}(c) $\la_2$ varies and
$(\la_1,\la_3,\la_4)$ remain constant, whereas in Fig.~\ref{fig4}(d)
$\la_3$ varies and $(\la_1,\la_2,\la_4)$ remain constant. We thus
arrive at the following description of these solutions:

\smallskip

$\bullet$ In the case of Fig.~\ref{fig4}(c) where
$\la_+^L=\la_+^R$, $\la_-^L>\la_-^R$ we have
\begin{equation}
 \la_1=\la_-^R,\quad \la_3=\la_-^L,\quad \la_4=\la_+^L=\la_+^R,
\end{equation}
and $\la_2$ depends on $\xi$ according to the
equation
\begin{equation}
 v_2(\la_-^R,\la_2,\la_-^L,\la_+^L)=\xi.
\end{equation}
The resulting wave pattern is obtained by substitution of these values
of $(\la_1,\la_2,\la_3,\la_4)$ into (\ref{eq18}) and (\ref{eq12}).
The left, small amplitude, edge of the DSW propagates with the
velocity
\begin{equation}
s_-=2\la_+^R+\frac{(\la_+^L-\la_-^L)^2}{2(\la_+^L+\la_-^L-2\la_+^R)}.
\end{equation}
The right edge corresponds to the soliton limit, propagating
with the velocity
\begin{equation}
 s_+=\frac12(\la_-^R+2\la_-^L+\la_+^L).
\end{equation}

$\bullet$ In a similar way, in the case of Fig.~\ref{fig4}(d)
where $\la_+^L>\la_+^R$,
$\la_-^L=\la_-^R$ we have
\begin{equation}\label{sh1}
 \la_1=\la_-^L=\la_-^R,\quad \la_2=\la_+^R,\quad \la_4=\la_+^L,
\end{equation}
and the dependence of $\la_3$ on $\xi$ is determined by the implicit
equation
\begin{equation}\label{sh2}
 v_3(\la_-^R,\la_+^R,\la_3,\la_+^L)=\xi.
\end{equation}
Substitution of the values
of $\la_i$ resulting from \eqref{sh1} and \eqref{sh2}
into (\ref{eq23}) and then (\ref{eq12}) yields the oscillatory DSW
structure for the physical variables $u$ and $h$. The left edge of the DSW
corresponds
to the soliton limit and this soliton moves with
the velocity
\begin{equation}\label{equa51}
 s_-=\frac12(\la_-^R+2\la_+^R+\la_+^L).
\end{equation}
Its right edge corresponds to the small amplitude limit propagating
with the velocity
\begin{equation}\label{equa52}
 s_+=2\la_+^L+\frac{(\la_+^R-\la_-^L)^2}{2(\la_+^R+\la_-^R-2\la_+^L)}.
\end{equation}

\section{Classification of solutions of the Riemann problem}\label{sec.classif}

For a given choice of initial conditions (\ref{eq4}), the solution of
the Riemann problem consists of combinations of the key elements listed
in the preceding section: plateaus, rarefaction waves and dispersive
shocks.  It is important to notice that if a RW or a DSW matches with
a plateau at both its left and right edges, then these plateaus share
one of their (dispersionless) Riemann invariants. For example, in
Figs.~\ref{fig4}(a) and (d) we have $\la_-^L=\la_-^R$ and in
Figs.~\ref{fig4}(b,c) we have $\la_+^L=\la_+^R$.

Also, in the case of a DSW, despite the fact that the dynamics inside
the shock region is described by four Riemann invariants, two of them
coincide with the dispersionless invariants of one of the plateaus at
the edges of the shock. Hence one may say that the value of one of the
dispersionless Riemann invariants ($\lambda_-$ say, as in the case of
Fig. \ref{fig4}(d)) is ``transferred'' through the DSW, although, if
it were computed using formula \eqref{eq31a}, one would find that it
strongly oscillates inside the DSW region. The equality
$\la^L_-=\la^R_-$ (for the case of \ref{fig4}(d)) connects the
parameters of the flow at both sides of the dispersive shock, and in
this sense it plays a role similar to that of the Rankine-Hugoniot
condition in the theory of viscous shocks (see
Refs. \onlinecite{gm-84,el-05}).

Due to this property of the dispersionless Riemann invariants, the
points corresponding to the edges of the DSW (or of the RW) must lie
on one of the parabolas along which the value of the dispersionless
Riemann invariant remains constant. Hence, the dispersionless
parabolas of Fig.~\ref{fig1} are useful tools for the classification
of all possible solutions. One should keep in mind that 
the parabolic
arcs symbolize the different types of solutions: a
physically acceptable single-valued RW (Fig.~\ref{fig4}(a,b)) or a
formal multi-valued solution (Fig.~\ref{fig4}(c,d)) which should be
replaced by a DSW correctly treated within the Whitham approach.
After these preliminary remarks we can
proceed to the classification of the wave structures.

\begin{figure}
\begin{center}
\includegraphics[width=0.99\linewidth]{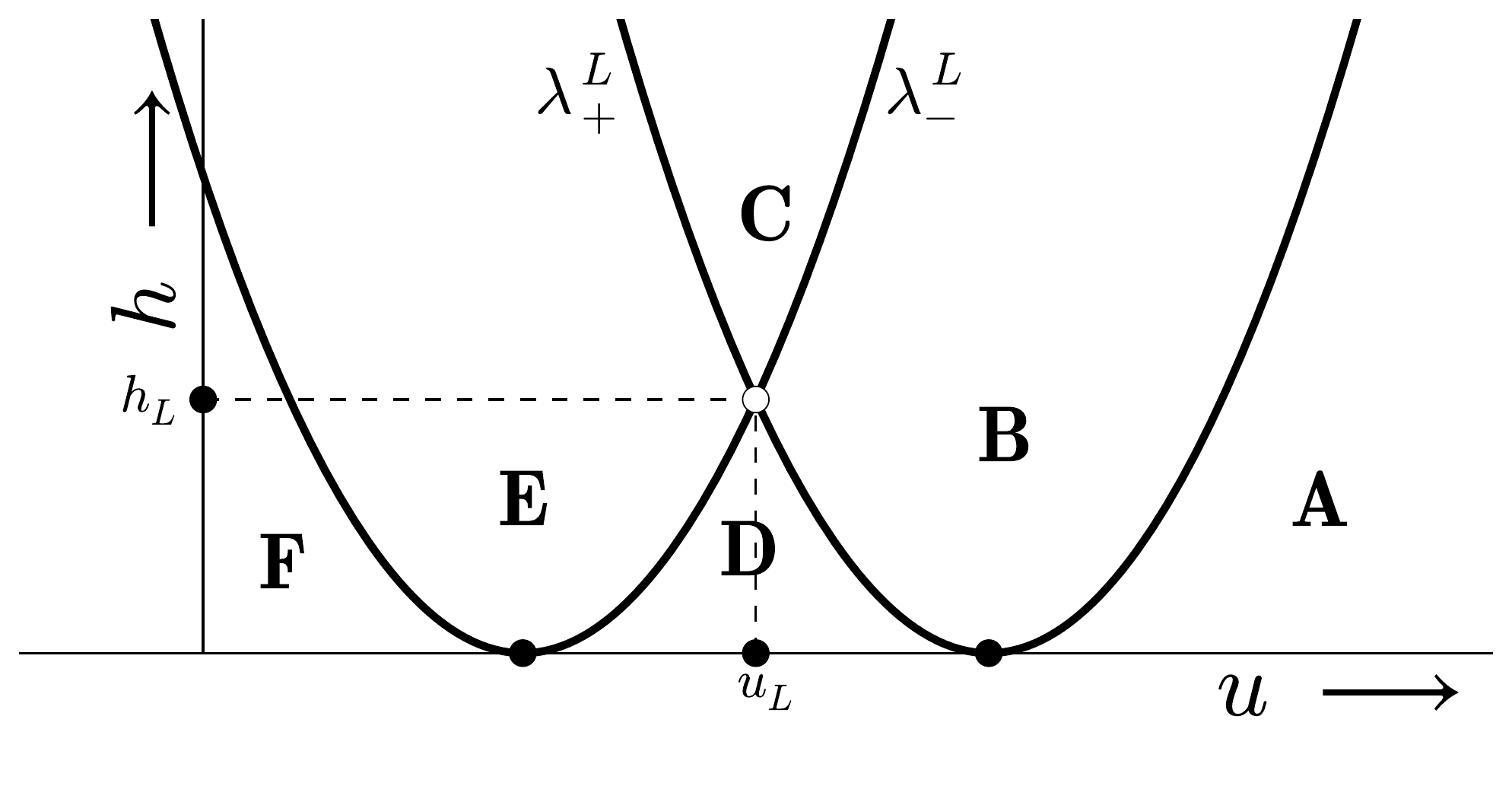}
\end{center}
\caption{Regions in the $(u,h)$ plane corresponding to different types of
  flow. The left boundary corresponds to point $L$ of coordinates
  $(u_L,h_L)$. The two parabolae are defined by the equations
  $h=(\tfrac12 u -\lambda_+^L)^2$ and $h=(\tfrac12 u
  -\lambda_-^L)^2$.  The type of flow depends on the region
(A,B, ..., or F) in which lies the right boundary point $R$ of
  coordinates $(u_R,h_R)$ .}
\label{fig5}
\end{figure}

The left and right boundaries of the whole wave structure connect with
undisturbed plateau regions whose parameters coincides with the
initial conditions (\ref{eq4}); for instance, in any situation one
should always have at the left boundary: $u(x/t\le s_-^L)=u_L$ and
$h(x/t\le s_-^L)=h_L$.  Consequently, the left and right edges
propagate into plateau regions represented by the two points
$(u_L,h_L)$ and $(u_R,h_R)$ in the $(u,h)$ plane. We represent in
Fig.~\ref{fig5} the two parabolas corresponding to the constant
dispersionless invariants $\la_{\pm}^L=u_L/2\pm\sqrt{h_L}$ including
their branches extending beyond the tangent points with the $u$-axis
(which  were represented as dashes lines in
Fig.~\ref{fig1}). These parabolas cut the physical half-plane $h>0$
into six domains labeled by the symbols $A,B,\ldots,F$.  Depending
on the domain in which the point $R$ with
coordinates $(u_R,h_R)$ lies, one has one of the six following
possible orderings of the left and right Riemann invariants:
\begin{equation}\label{eq14.1}
\begin{split}
& \mbox{A}:\; \la_-^L<\la_+^L<\la_-^R<\la_+^R,\;\;
\mbox{B}:\; \la_-^L<\la_-^R<\la_+^L<\la_+^R,\\
& \mbox{C}:\; \la_-^R<\la_-^L<\la_+^L<\la_+^R,\;\;
\mbox{D}: \la_-^L<\la_-^R<\la_+^R<\la_+^L,\\
& \mbox{E}:\; \la_-^R<\la_-^L<\la_+^R<\la_+^L,\;\;
 \mbox{F}:\; \la_-^R<\la_+^R<\la_-^L<\la_+^L.
\end{split}
\end{equation}
These six situations correspond to the six possible wave structures
resulting from the initial discontinuous profiles (\ref{eq4}).We shall
now describe their main properties and parameters. Note that, as
expected, the typology below does not depend on the absolute values of
$u_R$ and $u_L$, but only on their relative positions.

\smallskip

\begin{figure}
\begin{center}
\includegraphics[width=\linewidth]{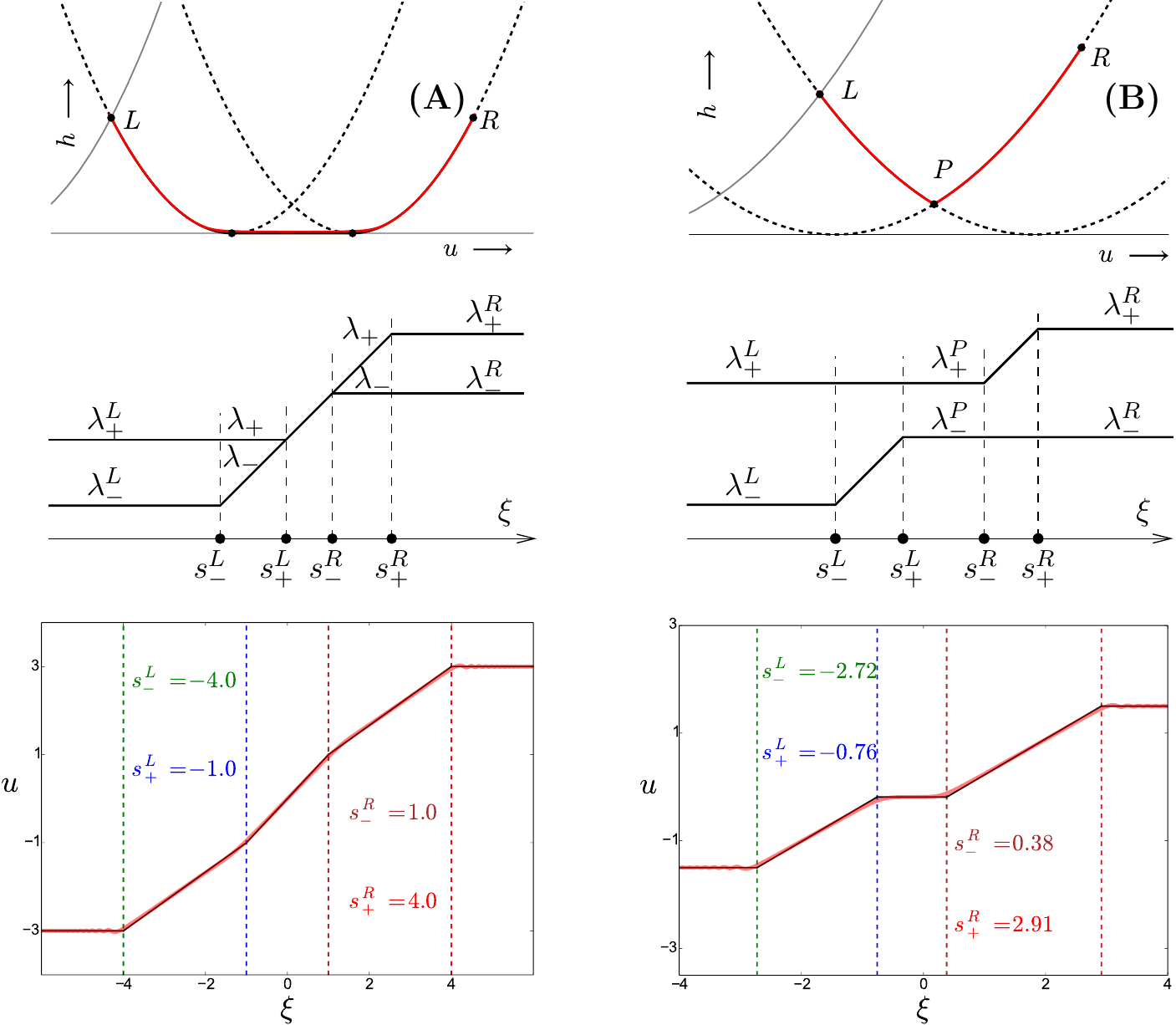}
\end{center}
\caption{Solutions in the cases (A) (three plots of the left column)
  and (B) (three plots of the right column). The initial profiles are
  characterized by $h_L=1$, $u_L=-3$, $h_R=1$, $u_R=3$ in case (A) and
  $h_L=1.5$, $u_L=-1.5$, $h_R=2$, $u_R=1.5$ in case (B). The upper
  plots display the behavior of the solution in the $(u,h)$ plane. The
  black solid line (almost perfectly recovered by the red line) is the
  result expected form the dispersionless approximation. The red solid
  line displays the results of numerical simulations.  The middle
  plots schematically represent the behavior of the Riemann invariant
  as functions of $\xi$. These are sketches, not on scale with the two
  other rows.  The lower plots compare the numerical simulations for
  the velocity field $u(\xi)$ (pink thick lines) with the analytic
  approach (black solid lines) from the dispersionless
  approximation. In these plots the vertical colored line are the
  velocities of the edges between the different components of the wave
  structure, as determined from \eqref{eq15.1} in case (A) and
  \eqref{eq15.3} in case (B).}
\label{figAetB}
\end{figure}

(A) In this case, the two rarefaction waves represented in
Fig.~\ref{fig2} (a) and (b) are combined into a single wave structure where
they are separated by an empty region (in which $h(x,t)=0$). The
velocities of the edges of the RWs are given by the formulae
\begin{equation}\label{eq15.1}
\begin{split}
&    s^L_-=u_L-\sqrt{h_L},\quad s^L_+=u_L+2\sqrt{h_L},\\
&    s^R_-=u_R-2\sqrt{h_R},\quad s^R_+=u_R+\sqrt{h_R}.
\end{split}
\end{equation}
The corresponding wave structure is displayed in
Fig.~\ref{figAetB}(A).  As expected, the dispersionless approximation
gives a very accurate description of the solution.

In the hydrodynamic context this situation corresponds to launching
two fluids in opposite directions with velocities so large that the
rarefaction waves are not able to fill the empty regions between them.

(B) Here the parabolas $u/2+\sqrt{h}=u_L/2+\sqrt{h_L}$ and
$u/2-\sqrt{h}=u_R/2-\sqrt{h_R}$ cross at the point $P=(u_P,h_P)$ with
the Riemann invariants $\la^P_{\pm}=u_P/2\pm\sqrt{h_P}$, and their
equality yields the values of the physical variables
\begin{equation}\label{eq15.2}
\begin{split}
&    u_P=\la^P_++\la^P_-=\frac12(u_L+u_R)+\sqrt{h_L}-\sqrt{h_R},\\
&
    h_P=\frac14(\la^P_+-\la^P_-)^2=
\frac14\left[\frac12(u_L-u_R)+\sqrt{h_L}+\sqrt{h_R}\right]^2\; .
\end{split}
\end{equation}
In this case, one has two rarefaction waves separated by a plateau
region which is represented by point $P$ in the $(u,h)$ plane of
Fig.~\ref{figAetB}(B). The velocities of the edges can be easily found
from the self-similar solutions (\ref{eq32a}) and \eqref{eq32b}:
\begin{equation}\label{eq15.3}
\begin{split}
    & s^L_-=v_-(\la^L_-,\la^L_+)=\frac12(3\la^L_-+\la^L_+)=u_L-\sqrt{h_L},\\
    & s^L_+=v_-(\la^P_-,\la^L_+)=\frac12(3\la^R_-+\la^L_+),\\
    & s^R_-=v_+(\la^P_-,\la^P_+)=\frac12(\la^R_-+3\la^L_+),\\
    & s^R_+=v_+(\la^R_-,\la^R_+)=\frac12(\la^R_-+3\la^R_+)=u_R+\sqrt{h_R}.
\end{split}
\end{equation}
The corresponding wave structure is displayed in
Fig.~\ref{figAetB}(B). As in case (A), the dispersionless approximation
gives a very accurate description of the solution.

Here the hydrodynamic interpretation is that
the two fluids are moving away from each other with velocities
lower than in the previous case (A), and
the rarefaction waves are now able to provide enough flux of fluid to create
a plateau in the region which separates them. This plateau has a fixed
value of the height $h$ and the flow velocity $u$.

\begin{figure}
\begin{center}
\includegraphics[width=\linewidth]{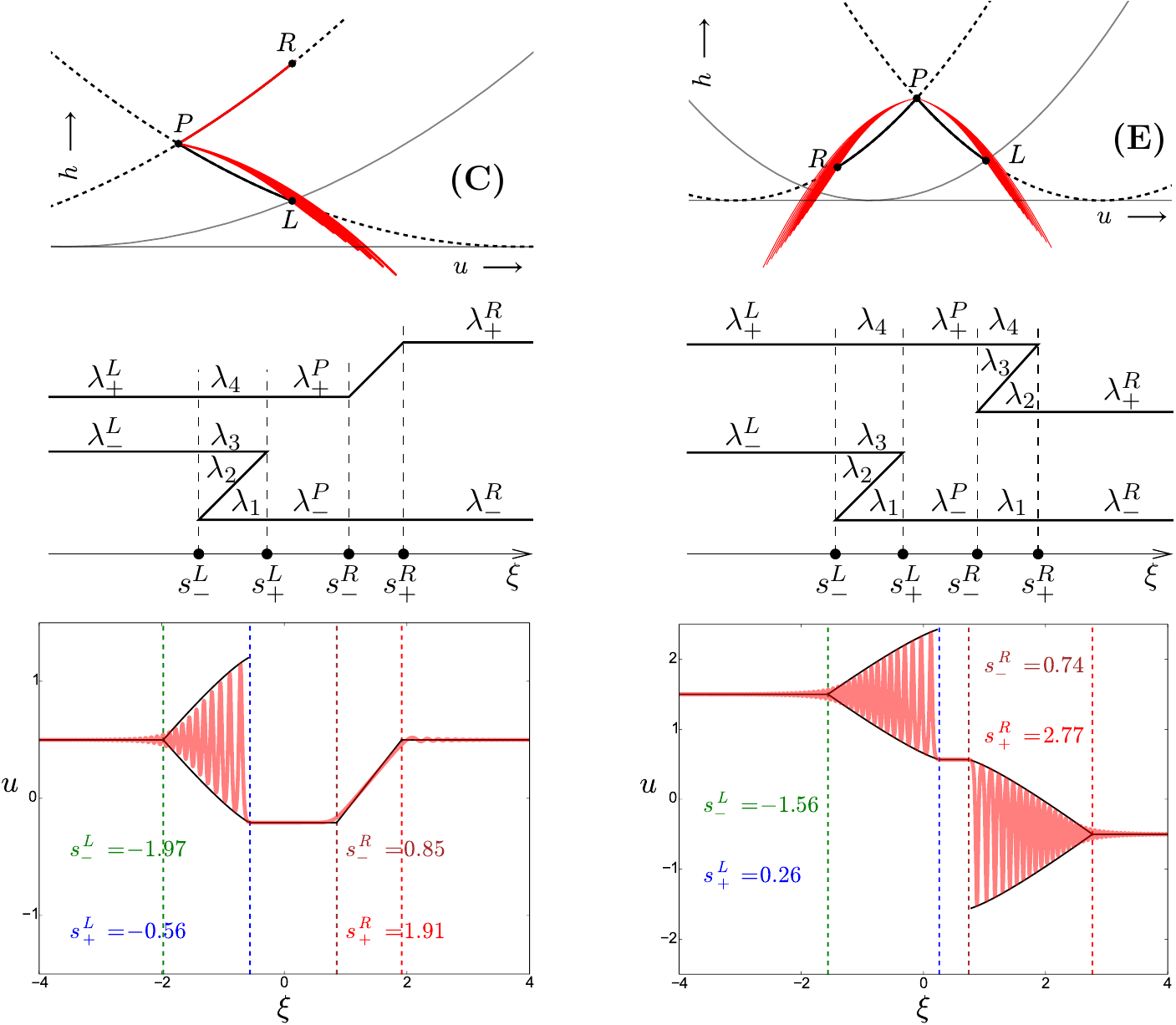}
\end{center}
\caption{Same as Fig. \ref{figAetB} for the cases (C) (three plots of
  the left column) and (E) (three plots of the right column). The
  initial profiles are characterized by $h_L=0.5$, $u_L=0.5$, $h_R=2$,
  $u_R=0.5$ in case (C) and $h_L=0.6$, $u_L=1.5$, $h_R=0.5$,
  $u_R=-0.5$ in case (E). The lower plots compares the numerical
  simulations for the velocity field $u(\xi)$ (pink thick lines) with
  the theoretical approach (black solid lines) composed of
  dispersionless approaches (in the plateau and the RW region) and
  Whitham modulation theory for the DSW. In the region of the DSW we
  only display the envelope of the nonlinear modulated wave. A more
  accurate comparison is done in Fig.~\ref{figCetE-h}.}
\label{figCetE}
\end{figure}

\smallskip

(C) In this case the initial profile evolves to form a DSW on the
left, a RW on the right and a plateau in between. The Riemann
invariants in the plateau region are $\la^P_-=\la^R_-$ and
$\la^P_+=\la^L_+$.  In the DSW region, the Riemann invariants behave
as schematically represented in Fig.~\ref{fig4}(c).  The edges of
the DSW propagate with velocities
\begin{equation}\label{eq15.4}
\begin{split}
& s^L_-=v_2(\la^R_-,\la^R_-,\la^L_-,\la^L_+)=
2\la_-^R+\frac{(\la^L_+-\la_-^L)^2}{2(\la^L_++\la^L_--2\la^R_-)},\\
& s^L_+=v_2(\la^R_-,\la^L_-,\la^L_-,\la^L_+)=\frac12(\la^R_-+2\la^L_-+\la^L_+),
\end{split}
\end{equation}
and the velocities of the edges of the RW are equal to
\begin{equation}\label{eq15.5}
    \begin{split}
    & s^R_-=v_+(\la^P_-,\la^P_+)=\frac12(\la^R_-+3\la^L_+),\\
    & s^R_+=v_+(\la^R_-,\la^R_+)=\frac12(\la^R_-+3\la^R_+).
    \end{split}
\end{equation}
This situation could be interpreted as if one fluid was colliding with
the other flowing away with such velocity that a plateau with
increased density is formed between them. The corresponding wave
structure is displayed in Fig.~\ref{figCetE}(C). The right RW and the
plateau region are correctly described by the dispersionless
approximation, as can be check on the upper plot of this figure where
the two approaches perfectly match between points $P$ (plateau region)
and $R$ (right boundary). Of course this is not true for the DSW: at
variance with the behavior expected on the basis of the dispersionless
approximation (black solid line), the numerical results (red solid
line) display large oscillations between points $P$ and $L$. This
behavior is, however, quite successfully described by the Whitham
approach, as can be seen in the lower plot of Fig.~\ref{figCetE}(C).

\smallskip

(D) Here we have a RW on the left and a DSW on the right with a plateau
in between. The Riemann invariants in the plateau region are equal
again to $\la^P_-=\la^R_-$, $\la^P_+=\la^L_+$. The velocities of the RW's
edges are equal to
\begin{equation}\label{eq15.6}
    \begin{split}
    & s^L_-=v_-(\la^L_-,\la^L_+)=\frac12(3\la^L_-+\la^L_+),\\
    & s^L_+=v_-(\la^R_-,\la^L_+)=\frac12(3\la^R_-+\la^L_+).
    \end{split}
\end{equation}
The behavior of the two dispersionless Riemann invariants in the
region of the rarefaction wave corresponds to the case illustrated in
Fig.~\ref{fig4}(b).  In the DSW region there are four
Riemann invariants which behave as schematically represented in
Fig.~\ref{fig4}(d), and the edges of the DSW propagate with velocities
\begin{equation}\label{eq15.7}
    \begin{split}
    & s^R_-=v_3(\la^R_-,\la^R_+,\la^R_+,\la^L_+)
=\frac12(\la^R_-+2\la^R_++\la^L_+),\\
    & s^R_+=v_3(\la^R_-,\la^R_+,\la^L_+,\la^L_+)=
    2\la_+^L+\frac{(\la^R_+-\la_-^R)^2}{2(\la^R_++\la^R_--2\la^L_+)}.
    \end{split}
\end{equation}
This situation is similar to the preceding one upon exchanging the roles
of the left and right fluids; we thus do not illustrate it by a figure.

\smallskip

(E) In this case the initial profile evolves
in two DSWs separated by a plateau, the parameters of
which are $\la^P_-=\la^R_-$ and $\la^P_+=\la^L_+$. The DSW's edges
propagate with velocities
\begin{equation}\label{eq15.8}
    \begin{split}
    & s^L_-=v_2(\la^R_-,\la^R_-,\la^L_-,\la^L_+)=
    2\la_-^R+\frac{(\la^L_+-\la_-^L)^2}{2(\la^L_++\la^L_--2\la^R_-)},\\
    & s^L_+=v_2(\la^R_-,\la^L_-,\la^L_-,\la^L_+)=
\frac12(\la^R_-+2\la^L_-+\la^L_+),\\
    & s^R_-=v_3(\la^R_-,\la^R_+,\la^R_+,\la^L_+)
=\frac12(\la^R_-+2\la^R_++\la^L_+),\\
    & s^R_+=v_3(\la^R_-,\la^R_+,\la^L_+,\la^L_+)=
    2\la_+^L+\frac{(\la^R_+-\la_-^R)^2}{2(\la^R_++\la^R_--2\la^L_+)}.
    \end{split}
\end{equation}
Here we have a collision of two fluids with `moderate' velocities: the
two DSWs do not overlap, but a central plateau region of increased
height is formed. This situation is represented in
Fig.~\ref{figCetE}(E). Again, the theoretical approach quite
accurately describes the numerical results (cf. the bottom row).

\begin{figure}[h]
\begin{center}
\includegraphics[width=\linewidth]{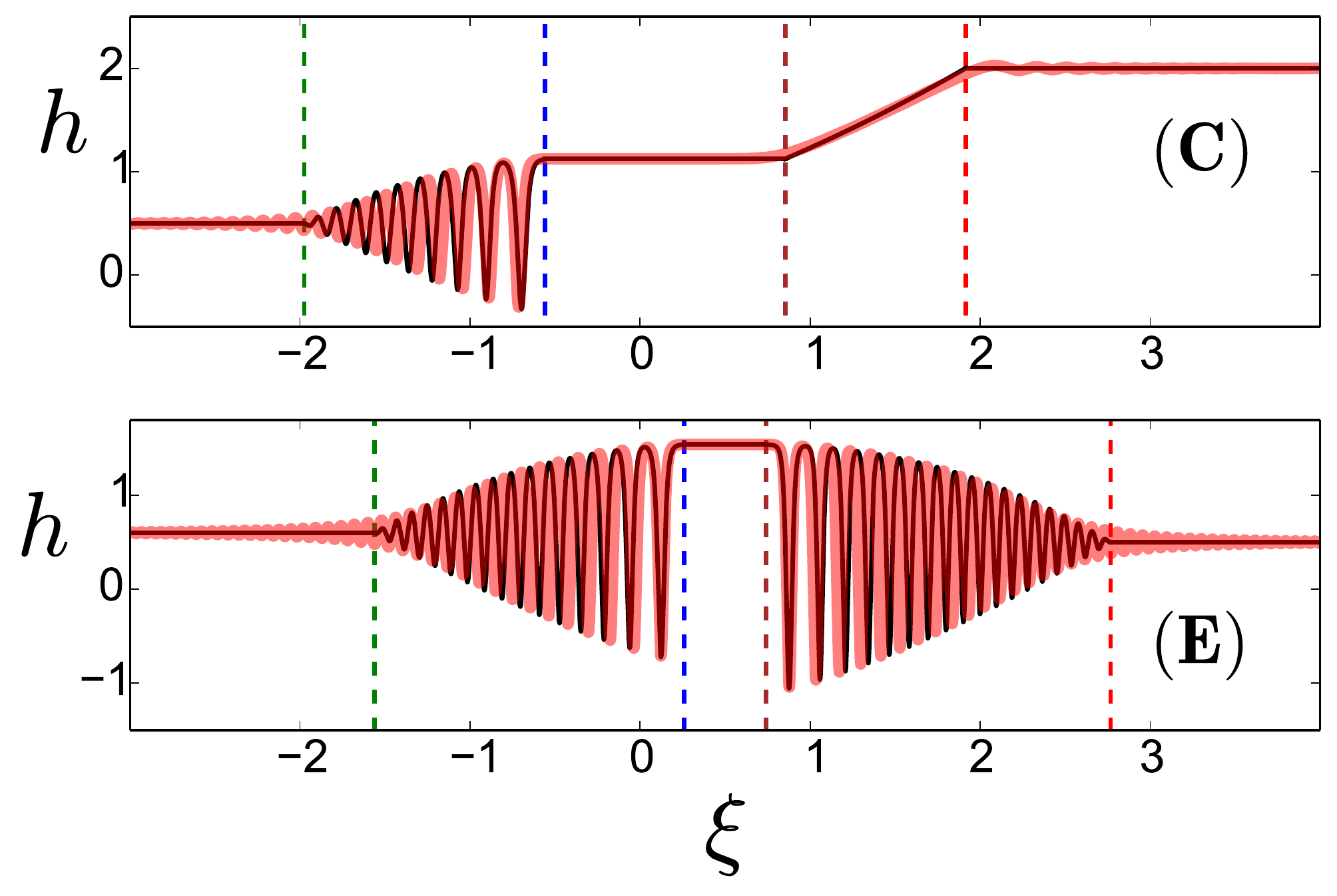}
\end{center}
\caption{$h$ as a function of $\xi$ for the same configurations as the
  ones depicted in Fig. \ref{figCetE}(C) and (E). The pink solid line
  represents the results of the numerical simulations and the black
  solid line is the theoretical result. Note the numerous excursions
  of $h(x,t)$ below zero.}
\label{figCetE-h}
\end{figure}

The upper part of the figure illustrates a phenomenon already present
in case (C): the large nonlinear oscillations in the DSW regions are
associated with locally negative values of $h(x,t)$. This phenomenon
is clearly seen in Fig.~\ref{figCetE-h}, which represents $h$ as a
function of $\xi$ for the two configurations (C) and (E) considered in
Fig.~\ref{figCetE}.  Although extended regions of constant and
negative values of $h$ lead to a dynamical instability (as clearly
seen from the dispersion relation \eqref{eq2}), nothing forbids local
excursions of $h$ below 0, and this is confirmed by the excellent
agreement of the numerical and theoretical results presented in
Fig.~\ref{figCetE-h}. Of course, in this case, the interpretation of
$h$ as being the height of a fluid surface becomes meaningless, but,
as explained in the introduction, the physical model behind the
nonlinear equations \eqref{eq3} can have an origin different from
shallow water physics.

\smallskip

(F) In this configuration the two fluids collide with velocities so
large that the central plateau observed in case (E) disappears: the DSWs
overlap and, on the basis of a similar situation observed for the
nonlinear Schr\"odinger equation \cite{eggk-95} and for the
Landau-Lifshitz equation \cite{Iva17}, one would expect that the
plateau is replaced by a new structure, separating two partial DSWs,
which can be approximated by a non-modulated cnoidal wave
(whose four Riemann invariants are constant) or more accurately by a
two-phase nonlinear wave.

Our numerical simulations show that this is not the case:
the collision of the DSW is here associated with a numerical
instability which we attribute to a physical dynamical instability of the
region of the overlapping DSWs.

\section{Dam break and piston problem}

In view of the particularities noticed in the precedent section --
possible negative values of $h$ and dynamical instabilities -- it is
interesting to study in more detail two model cases illustrating the
specificity of the Kaup-Boussinesq system. The first one is the dam
break problem which corresponds to a particular case of initial
conditions \eqref{eq4}. The second case is the piston problem. It
does not pertain to the same class of initial conditions, but
nevertheless provides an instructive insight on
non-modulated cnoidal waves whose stability
is questioned by the results obtained in case (F).

\subsection{Dam break problem}

This is the case where a semi-infinite constant height of water
expands into empty space, which would be a model of flow after the
abrupt breaking of a dam. Such a configuration is schematically
described by an initial condition of type \eqref{eq4} with
\begin{equation}\label{db1}
h_R=0 \quad\mbox{and}\quad u_R=0\; .
\end{equation}
On the basis of physical intuition, one expects that the time
evolution of this initial profile will result in a rarefaction wave
expanding into vacuum, as for the case illustrated in
Fig. \ref{fig2}(a). This is not quite correct: such a situation is
only reached when $u_L$ is sufficiently negative. More specifically,
the initial condition \eqref{db1} pertains either to case (A) when
$u_L<-2\sqrt{h_L}$, either to case (D) when $|u_L|<2\sqrt{h_L}$,
either to case (F) when $u_L>2\sqrt{h_L}$. In other words, this is
only when the initial left velocity (the initial velocity of the water
of the dam) is negative enough that a rarefaction wave is
observed. This behavior is different from the one observed in the
similar case for the nonlinear Schr\"odinger equation
\cite{eggk-95}. In the present case, the most natural situation where
$u_L=0$ (the water in the dam is initially steady) pertains to case
(D) for which the dam break leads to a DSW where the field $h$ becomes
negative. Only when $u_L$ becomes negative enough (in practice, when
it becomes lower than $-2\sqrt{h_L}$, i.e., when one reaches the
regime (A)) does the excursion of $h(x,t)$ below zero disappear.

\begin{figure}
\begin{center}
\includegraphics[width=\linewidth]{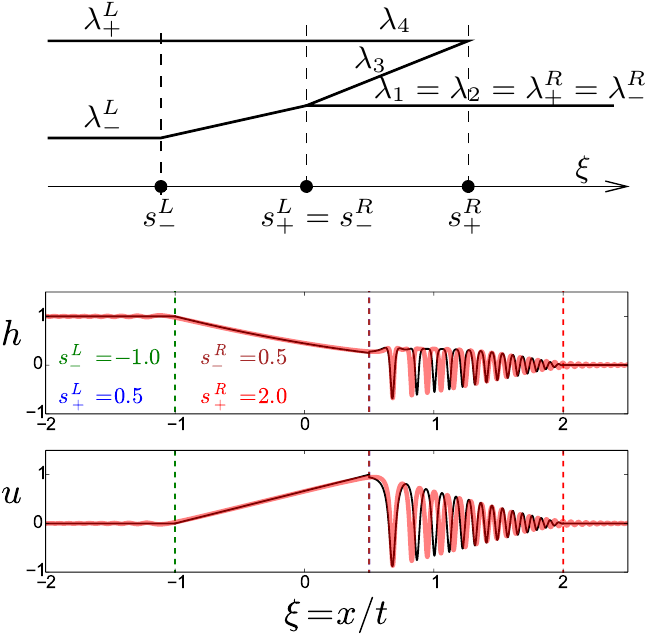}
\end{center}
\caption{The upper plot is a sketch of the behavior of the Riemann
  invariants describing the  dam break problem when
  $|u_L|<2\sqrt{h_L}$ (type (D) configuration). The dispersionless
  Riemann invariants describing the boundary conditions at the right
  are $\la_+^R=\la_-^R=0$, and at the left, $\la_\pm^L=\tfrac12
  u_L\pm\sqrt{h_L}$. The two lower plots compare the results of the
  numerical simulations (pink solid lines) with the theoretical
  results (black solid lines) for the boundary condition $h_L=1$ and
  $u_L=0$.}
\label{figDBP}
\end{figure}

This point deserves a slightly more detailed discussion: for the dam break
problem in case (D) one can easily check than the plateau region has a
vanishing extension ($s_+^L=\tfrac12 \la_+^L=s_-^R$). The behavior of
the Riemann invariants is depicted in Fig. \ref{figDBP}.
In this case $\la_1=\la_2$ and the DSW is described by a nonlinear
trigonometric wave \cite{gardner-2012,ckp-16} whose large amplitude
boundary (at $s_-^R$) corresponds to an algebraic soliton of type
\eqref{alg-sol} for which the largest value of $\mu$ is
$\la_4=\la_+^L$. Then, from \eqref{eq12}, the corresponding extremal
value of $h$ is $-\tfrac12 \la_4^2=-\tfrac12(\tfrac12 u_L +
\sqrt{h_L})^2$: it is always negative, and only vanishes when one
leaves regime (D) to enter regime (A), i.e., when $u_L\le
-2\sqrt{h_L}$: in this case the plateau $P$ and the right boundary $R$
coincide and the flow is of the type exemplified in
Fig. \ref{fig2}(a), which corresponds to a RW expanding into empty
space, as intuitively expected.

\subsection{Piston problem}

The piston problem corresponds to the situation where a hard wall (the
piston) is moving (in the case considered here, with a constant
positive velocity $V$) with respect to a steady fluid. We work
henceforth in the rest frame of the piston. In this frame the piston
is located at $x=0$, the fluid is incoming from the right with a
constant velocity $u_R=-V$ and a fixed constant depth $h_R$. The
boundary condition on the piston is $u(0,t)=0$: the fluid in contact
with the piston is at rest with respect to it. The boundary condition for the
height is taken as $h(0,t)=0$, or $h_x(0,t)=0$. These two conditions,
of Dirichlet or Neumann type, are equivalent if treated within the
Whitham approach, since the corresponding profiles differ only locally
near $0$, over characteristic lengths of order 1.

\begin{figure}
\begin{center}
\includegraphics[width=\linewidth]{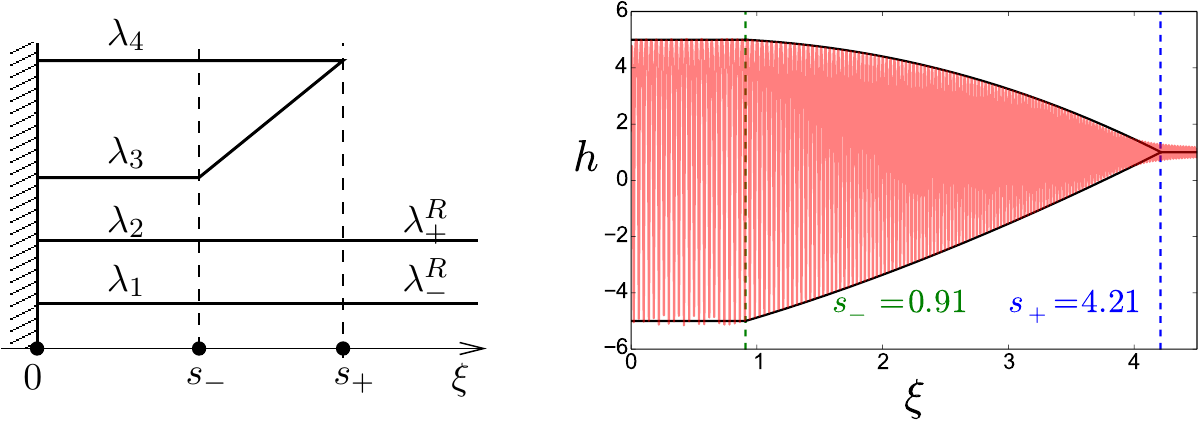}
\end{center} \caption{The piston problem for a velocity
  $V>2\sqrt{h_R}$. The left plot displays a sketch of the
  corresponding arrangement of Riemann invariants. In the rest frame
  of the piston the SCW occupies the region $0<\xi< s_-$, and the DSW
  the region $s_-<\xi<s_+$. The right plot compares the results of our
  numerical simulations for $h(x,t)$ with the prediction of Whitham
  theory. The boundary condition are $V=2.5=-u_R$ and $h_R=1$. In the
  numerical approach, the piston is mimicked by a strong repulsive
  potential.}\label{piston-SCW}
\end{figure}

For intermediate velocities $V$, the profile is of the type
characterized by the arrangement of Riemann invariants displayed in
Fig. \ref{fig4}(d): there is a plateau in contact with the piston,
then, at its right, a DSW, and finally a plateau corresponding to the
right boundary condition, characterized by $\la_{\pm}^R=\tfrac12
u_R\pm \sqrt{h_R}$. The plateau in contact with the piston is
characterized by a height $h_L$ (unknown at this point) and a velocity
$v_L=0$, hence $\lambda_\pm^L=\pm \sqrt{h_L}$. The constancy of the
lower Riemann invariant across the structure of Fig.~\ref{fig4}(d)
yields $\la_-^L=\la_-^R$ which fixes the value of
$h_L=(\sqrt{h_R}-u_R/2)^2$. The velocities of the edges of the DSW are
determined from \eqref{equa51} and \eqref{equa52}. In particular, the
boundary between the left plateau and the DSW has a velocity
$s_-=\sqrt{h_R}-V/2$. This velocity vanishes when $V=2\sqrt{h_R}$. For
piston velocities $V$ larger than this threshold, the plateau in
contact with the piston disappears and the structure of the flow
changes: in a good approximation it is represented by a stationary,
non-modulated cnoidal wave (SCW) in contact with the piston. This SCW
is connected to its right with a partial DSW, itself connecting to a
plateau defined by the right boundary condition. This corresponds to
the arrangement of Riemann invariants displayed in
Fig.~\ref{piston-SCW}.

The fact that the cnoidal wave located in the region $0< \xi< s_-$ in
Fig.~\ref{piston-SCW} is stationary reflects in the relation
$\la_1+\la_2+\la_3+\la_4=0$ which fixes its phase velocity to
zero. This yields
\begin{equation}\label{piston1}
\la_3+\la_4=-\la_1-\la_2=-u_R=V\; .
\end{equation}
Another condition is obtained by imposing that there is no flux
through the piston, therefore the average flux vanishes: $\langle h
u\rangle =0$ (it is evident that there is no contribution of the last term
in the first Eq.~(\ref{eq3}) to the average flux, since $\langle
u_{xx}\rangle\equiv0$ by virtue of local periodicity). The relation
\eqref{piston1} yields $s_1=0$ and Eqs.~\eqref{eq12} now read
$u=-2\mu$ and $h(\mu)=-s_2-2\mu^2$ where
$s_2=\la_1\la_2-(\la_1+\la_2)^2+\la_3\la_4$. Hence we get the
condition
\begin{equation}\label{piston2}
s_2\langle \mu\rangle + 2 \langle \mu^3\rangle =0\; ,
\end{equation}
where $\langle \mu^n\rangle=L^{-1}\oint \tfrac 12\mu^n {\rm
  d}\mu/\sqrt{-P(\mu)}$.  Condition \eqref{piston2} is fulfilled for
$s_3=(\la_1+\la_2)(\la_3\la_4-\la_1\la_2)=0$. Together with
\eqref{piston1} this yields a system for the yet unknown quantities
$\la_3$ and $\la_4$. The obvious solutions are (taking into account
the ordering \eqref{eq10a}) $\la_4=-\la_1$ and $\la_3=-\la_2$.  Then,
in the SCW, the height $h$ oscillates between the two opposite values
$\pm 2V\sqrt{h_R}$ and the velocity $v$ between $-V\pm
2\sqrt{h_R}$. The velocities of the edges of the DSW are $s_-=
v_3(\la_1,\la_2,\la_3,\la_4)$ and $s_+ = v_3(\la_1,\la_2,\la_4,\la_4)=
2\sqrt{h_R}+V - h_R/(V+\sqrt{h_R})$. The wavelength of the large
amplitude edge of the DSW (at $\xi=s_-$) is given by
\begin{equation}\label{wavelength-piston}
  L=\frac{2K(m^*)}{V},\quad\text{where}\quad m^*=\frac{4h_R}{V^2}
\end{equation}
is the modulus \eqref{eq19} of the elliptic functions; $m^*<1$ in the
present case since $V>2\sqrt{h_R}$.  As is checked in
Fig.~\ref{piston-SCW}, these predictions are in excellent agreement
with the numerical simulations.

Two comments are in order here: first, the numerical simulations show
that the cnoidal wave is weakly modulated, and, as suggested in
Refs.~\onlinecite{egp-01,bikbaev}, it should be more accurately
described as a two phase solution.  However, as seen in
Fig.~\eqref{piston-SCW} the modulation is small, and the approximate
description of the structure as a SCW is quite accurate. Second, and
more important, we have here an example of dynamically stable
non-modulated cnoidal wave with large amplitude ($h$ oscillate between
$\pm 5$ in the SCW region of Fig.~\ref{piston-SCW}). This is quite
different from the situation observed in the preceding section (region
(F) in Fig.~\ref{fig5}) where an expected SCW resulting from the
collision of two DSWs has proven unstable.  This example shows that
the wave structures in the piston problem may have properties quite
different from those arising from the evolution of initial
discontinuities.

\section{Conclusion}

In this paper we have developed a full classification of the wave
patterns evolving dynamically from initial discontinuities according
to the Kaup-Boussinesq equation with positive dispersion. At variance
with the case of negative dispersion considered in
Ref.~\onlinecite{egp-01}, the classification used here
follows closely the one for the nonlinear Schr\"odinger equation
\cite{eggk-95}, although there are a number of technical differences
caused by the possible negative value of the ``height'' field and also
by different representations of dark and bright ``soliton trains'' and
corresponding changes of the Whitham modulation equations.
This common behavior of the positive dispersion
Kaup-Boussinesq and the nonlinear Schr\"odinger equation is related
to the common sign of dispersion in both equations and will be
clarified in a forthcoming publication \cite{Iva17}.

Our results can find applications as approximations of the dynamics of
polarization waves in two-component Bose-Einstein condensates
\cite{ckp-16} and of magnetic systems with easy-plane anisotropy
\cite{ish-17}. Work in this direction is in progress.

\section*{Acknowledgments}
We are grateful to M.~A.~Hoefer for useful discussions.
AMK thanks Laboratoire de Physique Th\'eorique et Mod\`eles Statistiques
(Universit\'e Paris-Sud, Orsay) where this work was started, for
kind hospitality. This work was supported by the Russian Foundation
for Basic Research (project no.~16-01-00398) and by the French ANR under
  grant n$^\circ$ ANR-15-CE30-0017 (Haralab project).

\end{document}